\documentclass[proof]{WileyASNA-v1}
\usepackage{hyperref}
\articletype{Article Type}%

\received{26 April 2016}
\revised{6 June 2016}
\accepted{6 June 2016}

\begin{document}

\title{The Influence of Magnetic Field Geometry in Neutron Stars Crustal Oscillations}

\author[1]{Gibran Henrique de Souza*}

\author[1]{Ernesto Kemp** }

\authormark{Gibran Henrique de Souza \textsc{et al}}

\address{\orgdiv{Instituto de F\'{i}sica Gleb Wataghin}, \orgname{Universidade Estadual de Campinas -- UNICAMP}, 13083-859, \orgaddress{\state{S\~{a}o Paulo}, \country{Brazil}}}



\corres{ \email{*gibranhsouza@gmail.com; **kemp@ifi.unicamp.br}}


\abstract{In this work, we have studied oscillations in the crust of a neutron star which magnetic field has both dipolar and toroidal components, the former extends from the stellar interior to the outer space and the later is confined inside the star radius. Our study is based on the solutions we have got for perturbations in the star fluid, confined to the crust thickness. Our results are compared to the frequencies observed in the \textit{Soft Gamma Repeaters} signals.}


\jnlcitation{\cname{%
\author{}, 
\author{}, 
\author{}, 
\author{}, and 
\author{}} (\cyear{}), 
\ctitle{}, \cjournal{}, \cvol{}.}


\maketitle


\section{Introduction}\label{sec1}

Magnetars are neutron stars powered by very strong magnetic fields, that can reach over $10^{15}G$, observations of soft gamma-ray repeaters (SGRs) seem to indicate that these objects are magnetars, with high magnetic fields and low rotation rates (\cite{Becker2009}). There were three giant flares associated with these objects observed so far with quasiperiodic oscillations (QPOs), within a range of different frequencies detected in the late-time tail of these events. Table \ref{tab1}summarizes the data extracted from references \cite{Sotani2007, Olausen2014}. We notice that he first event in Table \ref{tab1}(the oldest),  has only one observed frequency because of the lack of precision from the observation satellite at that time, but with later improvement of the technology more frequencies were detected.

\begin{center}
\begin{table*}[t]%
\caption{Observed giant flares from \textit{SGR}s and measured frequencies (Hz). Each frequency correspond to a different oscillatory mode identified during the long-tail phase of the signal.\label{tab1}}
\centering
\begin{tabular*}{500pt}{llllllllllllll}
\toprule
\textbf{Object} & \textbf{Date of detection}  & \multicolumn{12}{l}  {\textbf{Frequencies (Hz)}} \\
\midrule
SGR 0526-66 & March 5, 1979  & 43.5 \\
SGR 1900+14 & August 27, 1998  & 28 & 54 & 84 & 155 \\
SGR 1806-20 & December 27, 2004  & 17& 18&  22&  26&  29&  37& 56& 92.5& 112& 150& 625.5& 1837  \\

\bottomrule
\end{tabular*}
\end{table*}
\end{center}

\section{The fluid perturbation equation}\label{sec2}

In what follows we shall assume that the oscillations in the fluid composing the non rotating neutron star can be described within the framework of the General Relativity. The magnetic field will be handled within the ideal magneto-hydrodynamics (MHD assumptions, which means that there is no separation of charge currents flowing through the stellar interior, and the field permeates the entire star. The metric inside a stationary and spherically symmetric star is described as follows (\cite{Colaiuda2009}):

\begin{equation}
ds^{2}= -e^{\nu (r)}dt^{2} + e^{\lambda  (r)}dr^{2} + r^{2}(d\theta ^{2}+sin^{2}\theta d\phi ^{2}
). 
\end{equation}
In this metric the stress energy tensor is:
\begin{equation}
T^{\alpha \beta }=(\rho +p+H^{2})u^{\alpha }u^{\beta }+\Big(p+\frac{H^{\alpha }}{2}\Big)g^{\alpha \beta }-H^{\alpha }H^{\beta };
 \end{equation} 
where $p$ is the fluid pressure, $\rho $ is the total energy density, $u^{\alpha }$ is the
fluid 4-velocity and $H^{\alpha }=B^{\alpha }/4\pi $ is the local magnetic field. The components of the dipolar magnetic field are given by (\cite{Colaiuda2009}): 

\begin{eqnarray}
  B_{\mu }= \bigg( 0, \quad \frac{2e^{\frac{\lambda (r)}{2}}}{r^{2}}a_{1}(r)\cos\theta, \quad -e^{\frac{-\lambda (r)}{2}}a_{1}(r)_{,r} \sin \theta , \\
 -\zeta e^{\frac{-\nu (r)}{2}}a_{1}(r)\sin^{2}\theta   \bigg);\nonumber
  \end{eqnarray}
  
where $a_{1}(r)$ is the radial profile of the vector potential, given by the Grad-Shafranov equation:

 \begin{eqnarray} \label{eq:4}
 e^{-\lambda (r)}\frac{d^{2}a_{1}}{dr^{2}}+\frac{e^{-\lambda (r)}}{2}\left(\frac{d\nu(r) }{dr}-\frac{d\lambda(r) }{dr} \right)\frac{da_{1}}{dr}+\\
\left( \zeta ^{2}e^{-\nu (r)}-\frac{2}{r^{2}} \right)a_{1}=4\pi c_{0}(\rho(r) +p(r))r^{2}.\nonumber  
 \end{eqnarray}
 
In equation (\ref{eq:4}) $4\pi c_{0}(\rho(r) +p(r))r^{2}$ is the source term and $\zeta $ is the ratio between the toroidal and poloidal field components.
The linear perturbation equations can be obtained by manipulations of the linearized Euler and energy conservation equations (\cite{Sotani2007}):
\begin{equation}
  \delta \Big(\big\{\delta ^{i}_{\alpha }+u^{i}u_{\alpha }\big\}T^{\alpha \beta }_{;\beta }\Big)=0;
\end{equation}
 \begin{equation}
   \delta \Big(u_{\alpha }T^{\alpha \beta }_{;\beta }\Big)=0;
 \end{equation}
together with the perturbed induction equations
 \begin{equation}
   \delta \Big\{\big(u^{\alpha }H^{\beta }-u^{\beta }H^{\alpha }\big)_{\beta }\Big\}=0.
 \end{equation}
 
 Developing these set of equations with the appropriate boundary conditions we reach the final form for the fluid perturbation equation is, for $Y(t,r)=e^{i\omega t}Y(r)$:
 \begin{eqnarray}
 \bigg[\mu (r)+\frac{(1+2\lambda _{2})(a_{1})^{2}}{\pi r^{4}}\bigg]\textit{Y}(r)_{,rr}+\\ \nonumber
\bigg\{\bigg(\frac{4}{r}+\frac{\nu _{,r}-\lambda _{,r}}{2}\bigg)\mu (r)+\mu (r)_{,r}+ \\ \nonumber
  +\frac{(1+2\lambda _{2})(a_{1})}{\pi r^{4}}\bigg[\bigg(\frac{\nu _{,r}-\lambda _{,r}}{2}\bigg)a_{1}+2a_{1,r}\bigg] \bigg\}\textit{Y}(r)_{,r}+\\ \nonumber
\bigg\{\bigg[\bigg(\rho +P+ \frac{(1+2\lambda _{2})(a_{1})^{2}}{\pi r^{4}}\bigg)e^{\lambda }+ \nonumber \\
 -\frac{\lambda _{2}(a_{1,r})^{2}}{2\pi r^{2}} \bigg]\omega ^{2}e^{-\nu }-(\lambda_{2} -2)\bigg(\frac{\mu (r)e^{\lambda }}{r^{2}}- \frac{\lambda _{2}(a_{1,r})^{2}}{2\pi r^{4}}\bigg)+\nonumber \\
  +\frac{(2+5\lambda _{1})(a_{1})}{2\pi r^{4}}\bigg[\bigg(\frac{\nu _{,r}-\lambda _{,r}}{2}\bigg)a_{1,r}+a_{1,rr} \bigg] \bigg\}\textit{Y}(r)=0.\nonumber
\end{eqnarray}
Where $\lambda_{1} =l(l+1)$, $\lambda _{2}=\frac{-l(l+1)}{(2l-1)(2l+3)}$ and $\mu (r)$ is the shear modulus.
We impose  as boundary conditions ${Y}(r_e)={Y}(r_i)=0$, where  $r_e$ and $r_i$ are respectively the external and internal radius of the crust, i.e. the stellar surface and the base of the crust. These conditions confine the perturbations inside the crust which corresponds to approximately $10\%$ of the stellar radius.

\section{Magnetic Field Configurations}\label{sec3}
\begin{figure}[t]
	\centerline{\includegraphics[angle=270,width=100mm]{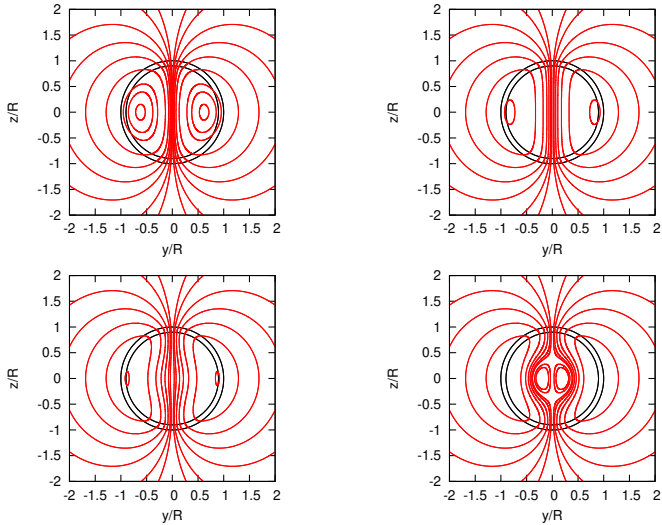}}
\caption{Magnetic field lines for typical configurations with a dipolar field and an increasing toroidal field component. The magnetic field is fixed as $B = 10^{15}G$ at the pole. The outer black circle indicates the surface of the star, while the inner circle indicates the base of the crust. The ratio $\zeta $ between the poloidal and toroidal components is 0.00 (top left), 0.26 (top right), 0.28 (bottom left) and 0.30 (bottom right).\label{fig1}}
\end{figure}

The magnetic field configurations (geometry) is the first step towards a better understanding of the impact over the structural properties of the neutron star. To determine the magnetic field properties we assumed a neutron star with mass of $M_{ns} = 1.40M_{\odot}$, with internal conditions described by different equations of state (EOS) (\cite{Douchin2001,Akmal1998,Lackey2006,Negele1973,Steiner2012}). Our results show, by inspection, that the magnetic field configuration have almost no dependence with the EOS. Thus, for illustration purposes we show our results for the SLy EOS (\cite{Douchin2001}) applied for both the core and the crust. We choose arbitrarily this EOS as the standard to show the magnetic field lines configuration (see Figure \ref{fig1}). For each $\zeta $ value we have a different field configuration, ranging from a pure dipolar field, $\zeta =0$, to a disjointed one, $\zeta =0.30$.

\section{Influence of the magnetic field amplitude on the frequencies}\label{sec4}
In the following we discuss the influence of the magnetic field amplitude on the frequencies for the fundamental mode, $n=0$, for the angular moment ranging from $l=2$, for a set of different core and crust equations of state \cite{Douchin2001,Akmal1998,Lackey2006,Negele1973,Steiner2012}. We choose these set of EOS to extent our studies to well known EOS in the literature.

\begin{figure}[t]
	\centerline{\includegraphics[width=78mm,height=13pc]{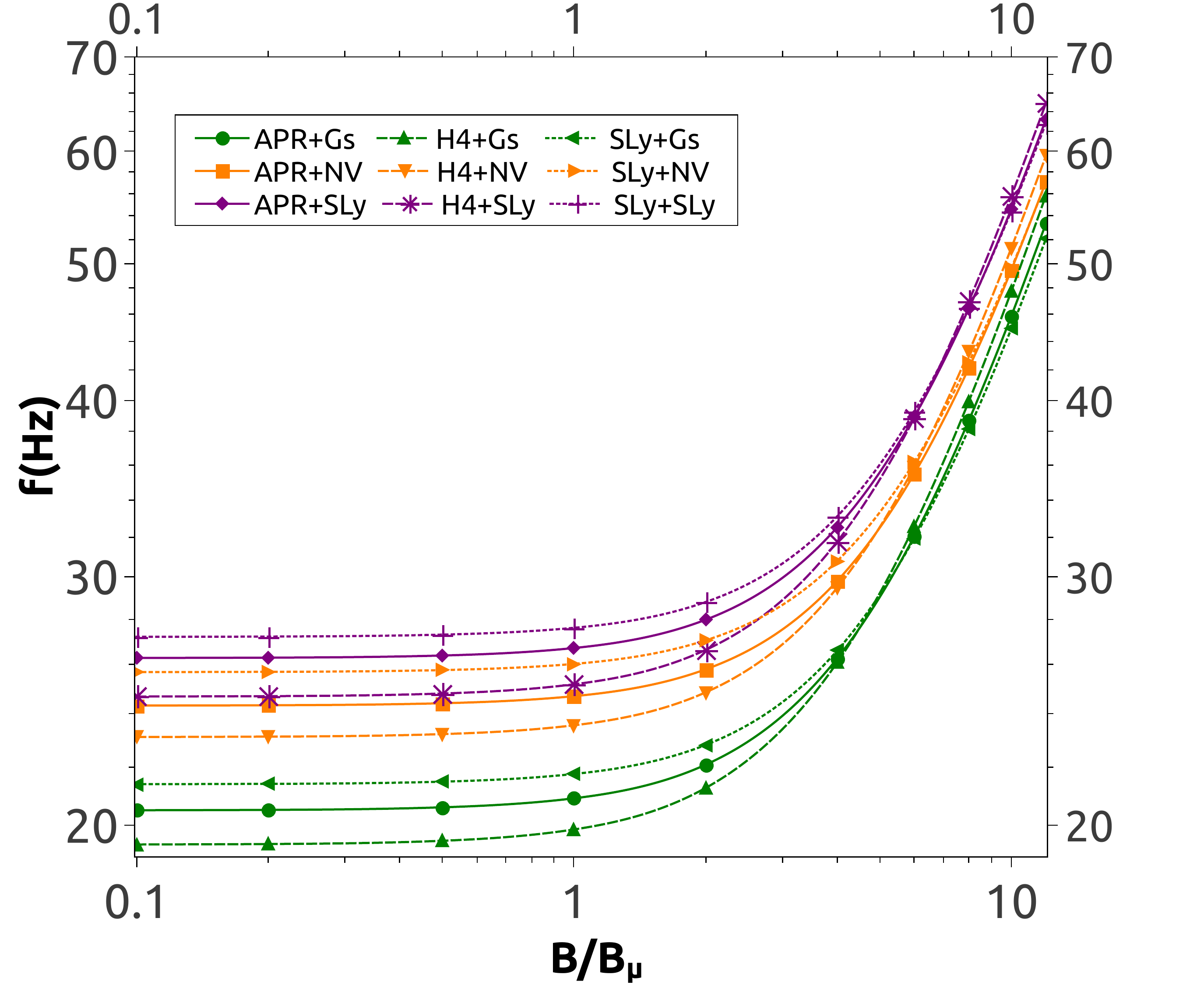}}
\caption{The crust frequency for the angular moment $l=2$. The field is normalized to $4 \times 10^{15}G$.\label{fig2}}
\end{figure}

We can approximate the general behavior of the magnetic field using the formula:
\begin{equation}
  \frac{f(B)}{f^{0}}\approx \bigg[1+\alpha \bigg(\frac{B}{4\times 10^{15}G} \bigg)^{2} \bigg]^{1/2},
\end{equation}
Where $f^{0}$ is the frequency in the absence of a magnetic field and $\alpha$ is a fitting constant. The quadratic monotonic behavior for all set of equations of state manifests the evidence for an universal behavior for all stars. The average lower frequencies are related to smaller magnetic field, and the opposite is also true.

The studies for the fundamental mode $n=0$, case $l=2$, showed also a general behavior for the fitting constant $_{2}\alpha_{0} $ for the set of equations of state in function of the field geometry parameter $\zeta^{2} $. 
\begin{figure}
  \centering
\centerline{\includegraphics[width=78mm,height=13pc]{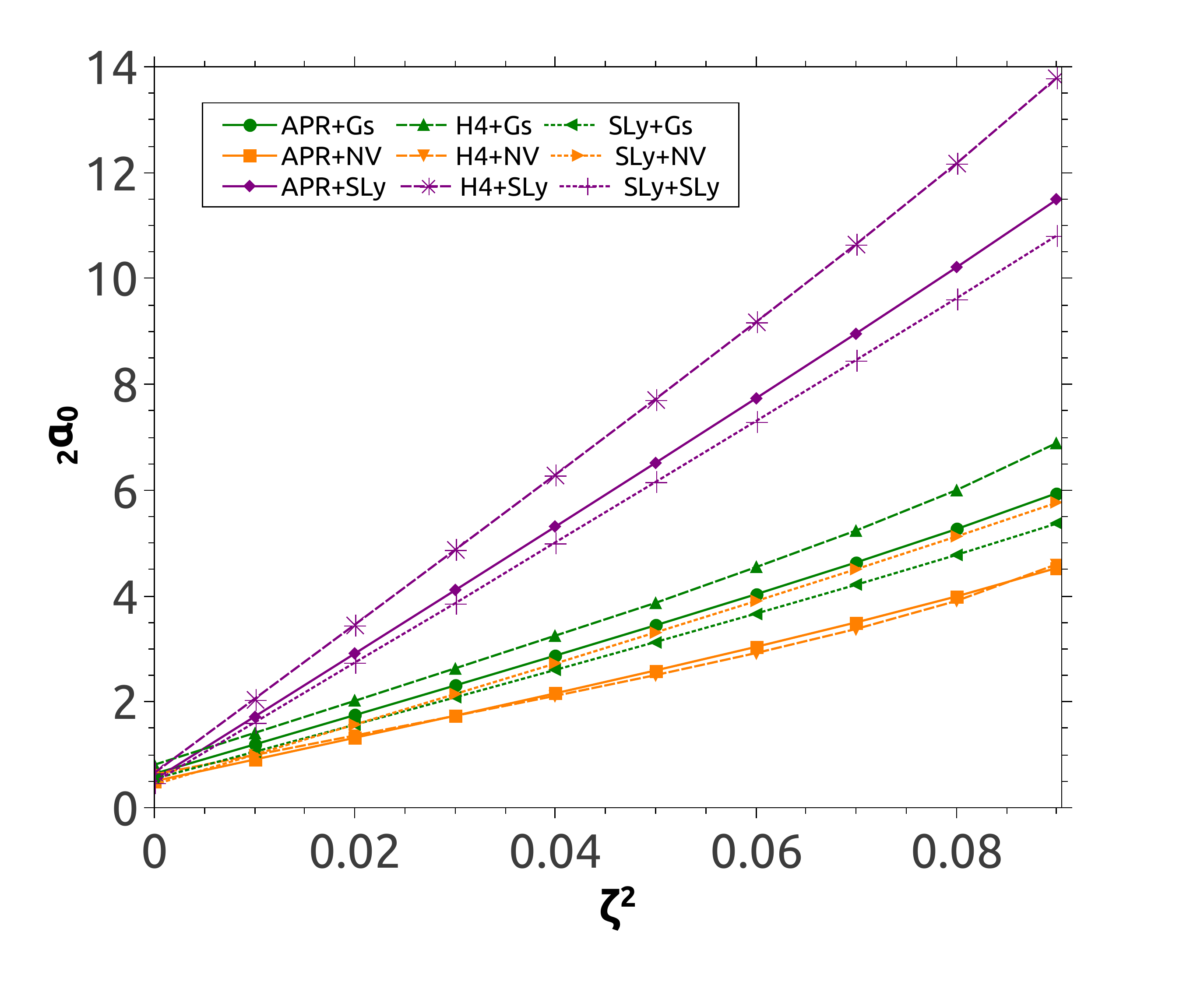}}
   \caption{The fitting constant $_{2}\alpha_{0} $ in function of the field parameter $\zeta^{2} $. We can see a quasi-linear behavior, the parameter $\zeta$ represents the ratio between the poloidal to toroidal field's components. As the toroidal component increases the frequencies become more sensible to the field amplitude.\label{fig3}}
   \end{figure}

\section{Influence of the magnetic field geometry on the frequencies}\label{sec5}

In the set of Figures $4-6$ we illustrate the influence of the magnetic field geometry on the frequencies for the fundamental mode, $n=0$. The angular momentum ranges from $l=2$ to $l=10$ and the field amplitude is fixed in $10^{15}G$, but the parameter $\zeta ^{2}$ is varying.

\begin{figure}[t]
	\centerline{\includegraphics[width=78mm,height=13pc]{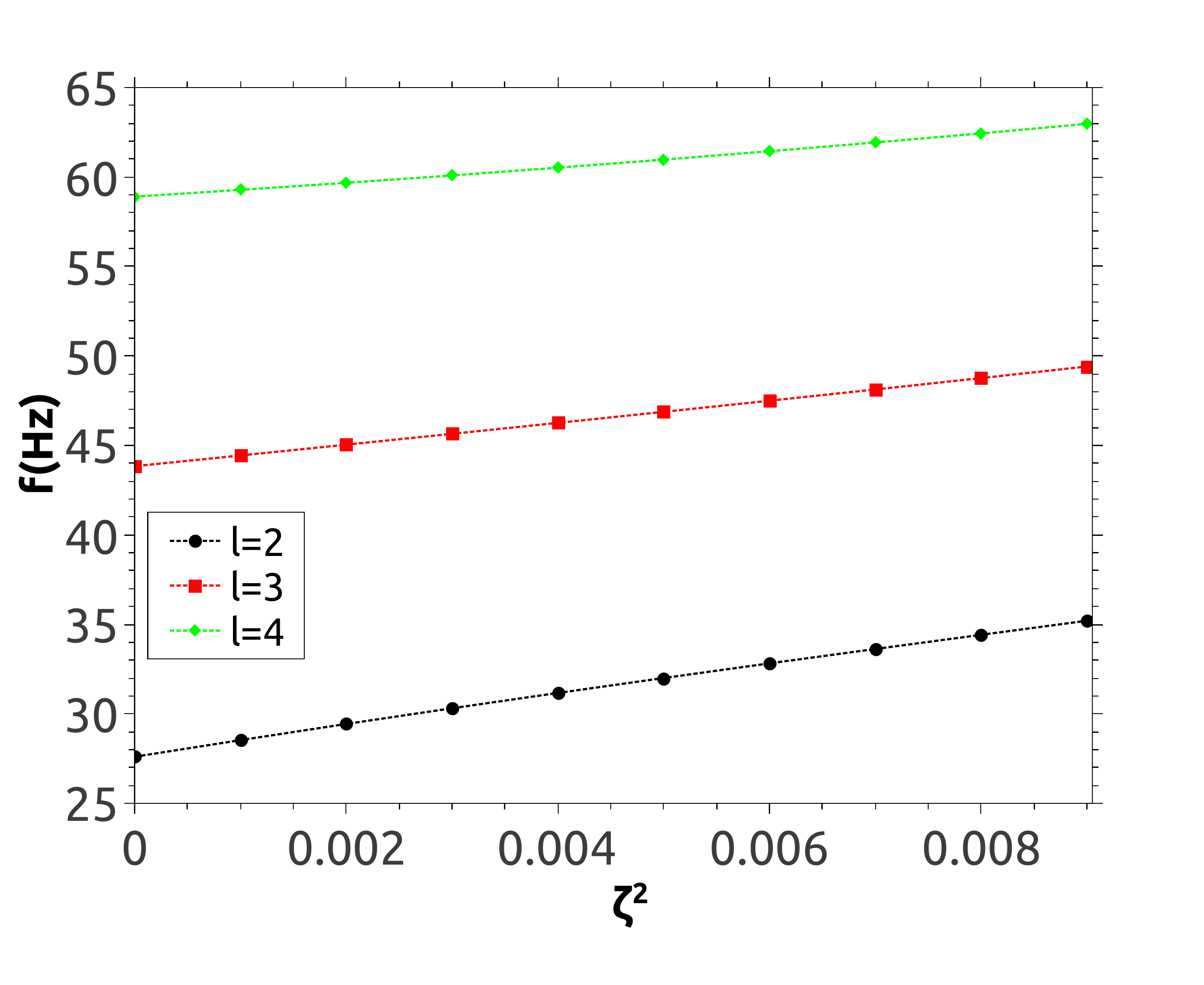}}
\caption{The crust frequency for the angular moments $l=2,3,4$. The field is fixed in $10^{15}G$ in both poles for all values of $\zeta $.\label{fig4}}
\end{figure}

\begin{figure}[t]
	\centerline{\includegraphics[width=78mm,height=13pc]{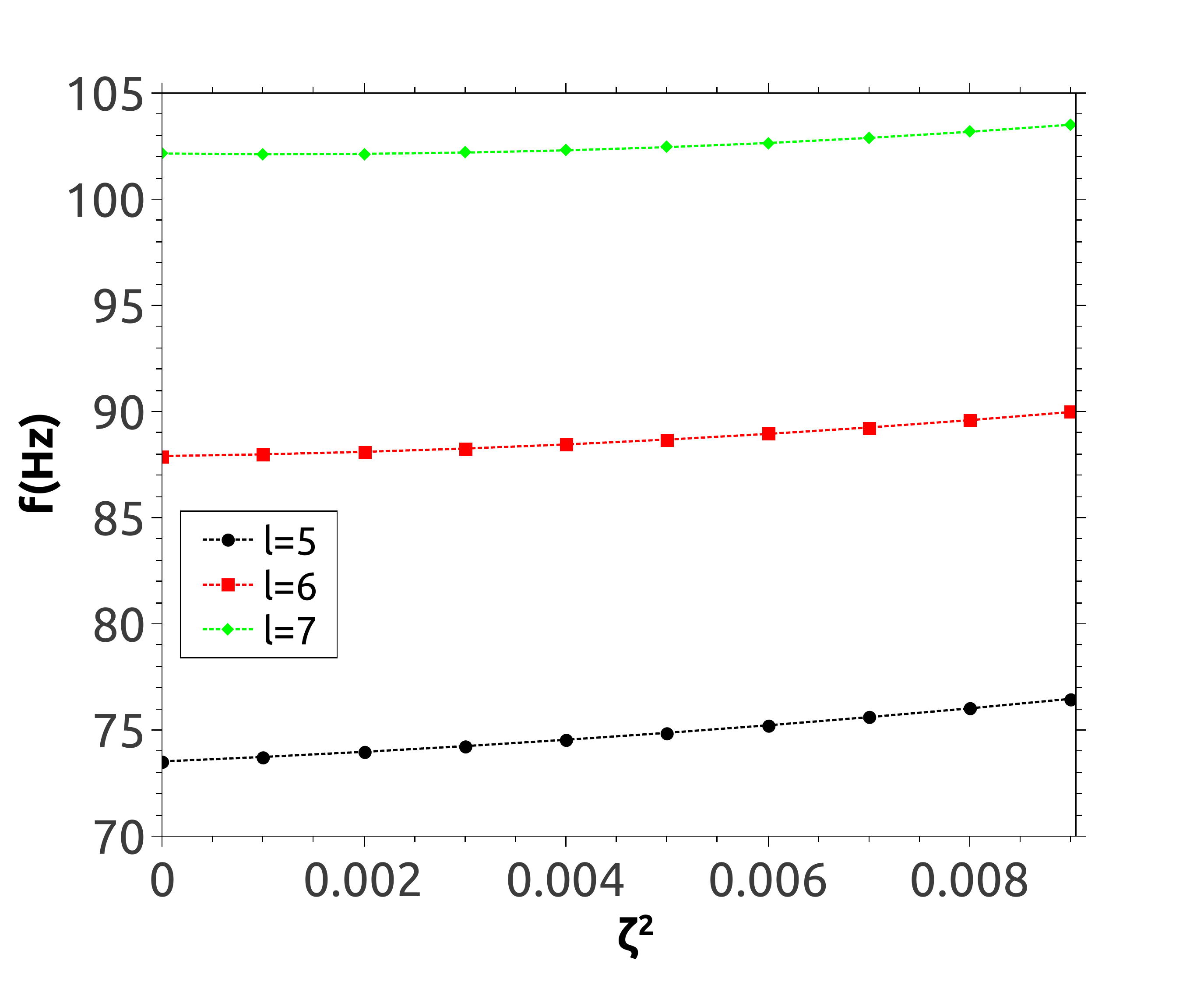}}
\caption{The crust frequency for the angular moments $l=2,3,4$. The field is fixed in $10^{15}G$ in both poles for all values of $\zeta $.\label{fig5}}
\end{figure}

\begin{figure}[t]
	\centerline{\includegraphics[width=78mm,height=11pc]{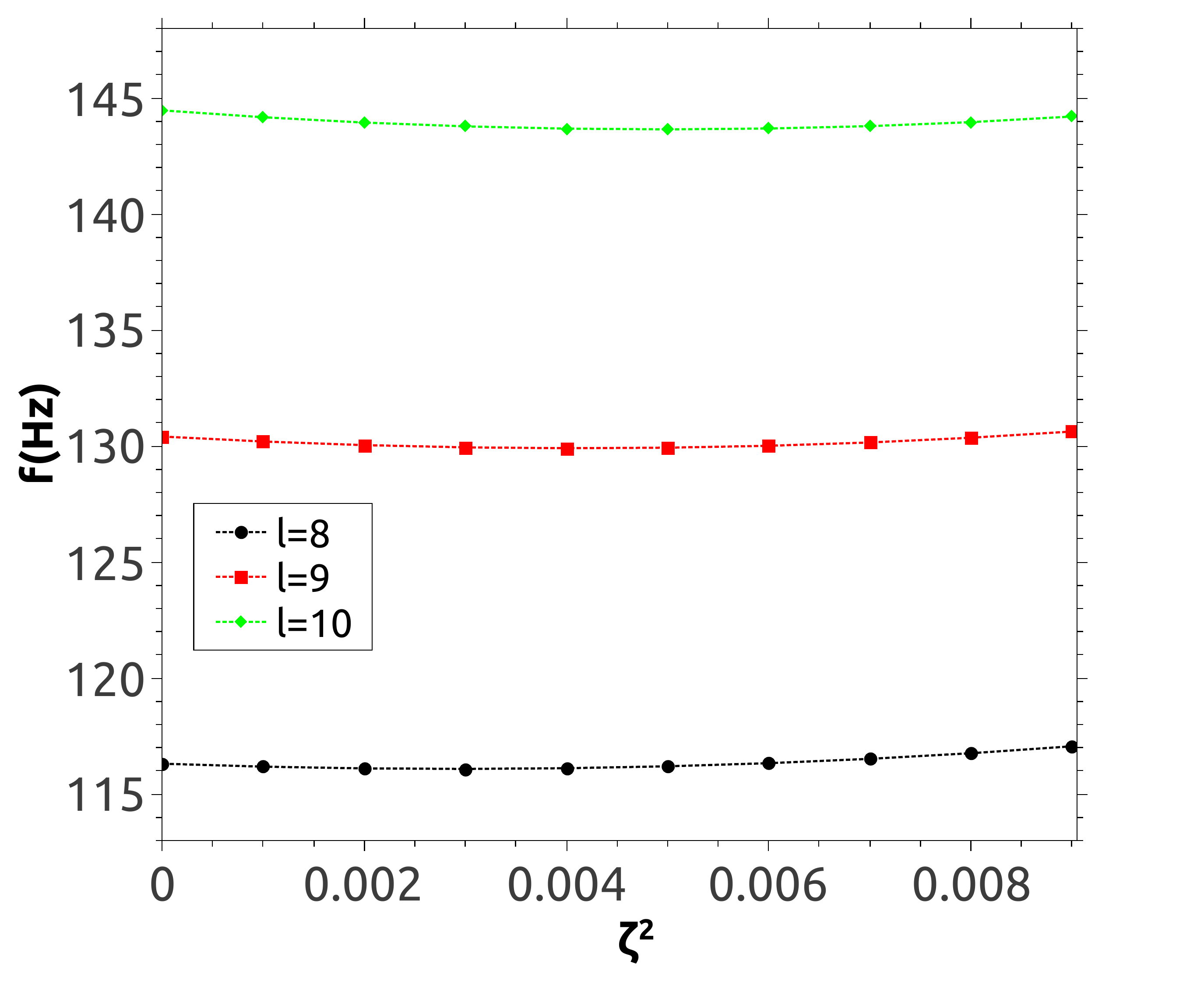}}
\caption{The crust frequency for the angular moments $l=8, 9, 10$. The field is fixed in $10^{15}G$ in both poles for all values of $\zeta $.\label{fig6}}
\end{figure}

Figure \ref{fig4}shows the behavior for $l=2,3,4$, highlighting the increasing monotonic feature, these angular moments are more sensible to the field geometry. Figure \ref{fig5}shows the behavior for $l=5,6,7$, highlighting the tendency to a slightly varying response frequencies, with only small variations, in comparison with figure 4. Finally, figure \ref{fig6}shows a stable behavior for frequencies $l=8,9,10$, for these angular moments the field geometry seems to play a negligible role on the crustal oscillations.

\section{Conclusions}\label{sec5}

We have presented here a careful analysis of the frequencies of torsional modes of oscillation of the magnetized neutron stars crust, motivated by the observations of QPOs in the tail of the observed giant flares of SGRs and the prospect of doing neutron star asteroseismology. We have studied the magnetic field configurations using different EOS and we found almost no dependence between them.  Then we explored the influence of the obtained configurations, whose geometry was encoded in the variable $\zeta$ , over the frequencies generated by perturbations in the crust matter. We found that the lower angular momenta are more influenced by the field geometry and could carry observable information to experiments. Nonetheless, bigger angular momenta can be used to study the field intensity. In the big picture, our results show that a pure dipolar magnetic field can not generate a large range of frequencies, requiring a small toroidal component restricted to the star's interior in order to explain the data in table 1. We should emphasize the toroidal component, as included by our approach, was not enough to fit all the data, indicating that a more detailed model for the core and crust coupling is necessary to explain all the observed frequencies that can be associated with the neutron star crust oscillations.

Our results show that only some field configurations can explain the core of the measured frequencies. Moreover, new studies claim that a magnetic field in neutron stars evolve with the time (\cite{Braithwaite2006}, \cite{Gabler2013}). The overlap of these facts suggests that only some specific field configurations can trigger the rare giant flares observed from SGRs.
\section*{Acknowledgments}
EK had financial support from the S\~{a}o Paulo Research Foundation - FAPESP (grant 2014/19164-6) to attend the STARS/SMFNS 2019.



\appendix


\bibliography{bibliography}%

\begin{thebibliography}{}

\bibitem [\protect \citeauthoryear {%
Akmal%
, Pandharipande%
\BCBL {}\ \BBA {} Ravenhall%
}{%
Akmal%
\ \protect \BOthers {.}}{%
{\protect \APACyear {1998}}%
}]{%
Akmal1998}
\APACinsertmetastar {%
Akmal1998}%
\begin{APACrefauthors}%
Akmal, A.%
, Pandharipande, V\BPBI R.%
\BCBL {}\ \BBA {} Ravenhall, D\BPBI G.%
\end{APACrefauthors}%
\unskip\
\newblock
\APACrefYearMonthDay{1998}{}{},
\newblock
\unskip
\newblock
\APACjournalVolNumPages{Phys. Rev. C}{58}{}{1804}.
\PrintBackRefs{\CurrentBib}

\bibitem [\protect \citeauthoryear {%
Becker%
}{%
Becker%
}{%
{\protect \APACyear {2009}}%
}]{%
Becker2009}
\APACinsertmetastar {%
Becker2009}%
\begin{APACrefauthors}%
Becker, W.%
\end{APACrefauthors}%
\unskip\
\newblock
\APACrefYear{2009},
\newblock
\APACrefbtitle {Astrophysics and Space Science Library} {Astrophysics and Space
  Science Library}\ (\PrintOrdinal{1}\ \BEd).
\newblock
\APACaddressPublisher{Berlin}{Springer}.
\PrintBackRefs{\CurrentBib}

\bibitem [\protect \citeauthoryear {%
Braithwaite%
\ \BBA {} Nordlund%
}{%
Braithwaite%
\ \BBA {} Nordlund%
}{%
{\protect \APACyear {2006}}%
}]{%
Braithwaite2006}
\APACinsertmetastar {%
Braithwaite2006}%
\begin{APACrefauthors}%
Braithwaite, J.%
\BCBT {}\ \BBA {} Nordlund, A.%
\end{APACrefauthors}%
\unskip\
\newblock
\APACrefYearMonthDay{2006}{}{},
\newblock
\unskip
\newblock
\APACjournalVolNumPages{A\&A}{450}{}{1077}.
\PrintBackRefs{\CurrentBib}

\bibitem [\protect \citeauthoryear {%
Colaiuda%
, Ferrari%
, Gualtieri%
\BCBL {}\ \BBA {} Pons%
}{%
Colaiuda%
\ \protect \BOthers {.}}{%
{\protect \APACyear {2008}}%
}]{%
Colaiuda2009}
\APACinsertmetastar {%
Colaiuda2009}%
\begin{APACrefauthors}%
Colaiuda, A.%
, Ferrari, V.%
, Gualtieri, L.%
\BCBL {}\ \BBA {} Pons, J\BPBI A.%
\end{APACrefauthors}%
\unskip\
\newblock
\APACrefYearMonthDay{2008}{}{},
\newblock
\unskip
\newblock
\APACjournalVolNumPages{Mon. Not. R. Astron. Soc.}{385}{}{2080}.
\PrintBackRefs{\CurrentBib}

\bibitem [\protect \citeauthoryear {%
Douchin%
\ \BBA {} Haensel%
}{%
Douchin%
\ \BBA {} Haensel%
}{%
{\protect \APACyear {2001}}%
}]{%
Douchin2001}
\APACinsertmetastar {%
Douchin2001}%
\begin{APACrefauthors}%
Douchin, F.%
\BCBT {}\ \BBA {} Haensel, P.%
\end{APACrefauthors}%
\unskip\
\newblock
\APACrefYearMonthDay{2001}{}{},
\newblock
\unskip
\newblock
\APACjournalVolNumPages{A\&A}{380}{}{151}.
\PrintBackRefs{\CurrentBib}

\bibitem [\protect \citeauthoryear {%
Gabler%
, Duran%
, Stergioulas%
, Font%
\BCBL {}\ \BBA {} Muller%
}{%
Gabler%
\ \protect \BOthers {.}}{%
{\protect \APACyear {2013}}%
}]{%
Gabler2013}
\APACinsertmetastar {%
Gabler2013}%
\begin{APACrefauthors}%
Gabler, M.%
, Duran, P\BPBI C.%
, Stergioulas, N.%
, Font, J\BPBI A.%
\BCBL {}\ \BBA {} Muller, E.%
\end{APACrefauthors}%
\unskip\
\newblock
\APACrefYearMonthDay{2013}{}{},
\newblock
\unskip
\newblock
\APACjournalVolNumPages{Mon. Not. R. Astron. Soc.}{430}{}{1811}.
\PrintBackRefs{\CurrentBib}

\bibitem [\protect \citeauthoryear {%
Lackey%
, Nayyar%
\BCBL {}\ \BBA {} Owen%
}{%
Lackey%
\ \protect \BOthers {.}}{%
{\protect \APACyear {2006}}%
}]{%
Lackey2006}
\APACinsertmetastar {%
Lackey2006}%
\begin{APACrefauthors}%
Lackey, B\BPBI D.%
, Nayyar, M.%
\BCBL {}\ \BBA {} Owen, B\BPBI J.%
\end{APACrefauthors}%
\unskip\
\newblock
\APACrefYearMonthDay{2006}{}{},
\newblock
\unskip
\newblock
\APACjournalVolNumPages{Phys. Rev. D}{73}{}{024021}.
\PrintBackRefs{\CurrentBib}

\bibitem [\protect \citeauthoryear {%
Negele%
\ \BBA {} Vautherin%
}{%
Negele%
\ \BBA {} Vautherin%
}{%
{\protect \APACyear {1973}}%
}]{%
Negele1973}
\APACinsertmetastar {%
Negele1973}%
\begin{APACrefauthors}%
Negele, J\BPBI W.%
\BCBT {}\ \BBA {} Vautherin, D.%
\end{APACrefauthors}%
\unskip\
\newblock
\APACrefYearMonthDay{1973}{}{},
\newblock
\unskip
\newblock
\APACjournalVolNumPages{Nuclear Physics A}{207}{}{298}.
\PrintBackRefs{\CurrentBib}

\bibitem [\protect \citeauthoryear {%
Olausen%
\ \BBA {} Kaspi%
}{%
Olausen%
\ \BBA {} Kaspi%
}{%
{\protect \APACyear {2014}}%
}]{%
Olausen2014}
\APACinsertmetastar {%
Olausen2014}%
\begin{APACrefauthors}%
Olausen, S\BPBI A.%
\BCBT {}\ \BBA {} Kaspi, V\BPBI M.%
\end{APACrefauthors}%
\unskip\
\newblock
\APACrefYearMonthDay{2014}{}{},
\newblock
\unskip
\newblock
\APACjournalVolNumPages{Astrophysical Journal Supplement Series}{212}{}{6}.
\PrintBackRefs{\CurrentBib}

\bibitem [\protect \citeauthoryear {%
Sotani%
, Kokkotas%
\BCBL {}\ \BBA {} Stergioulas%
}{%
Sotani%
\ \protect \BOthers {.}}{%
{\protect \APACyear {2007}}%
}]{%
Sotani2007}
\APACinsertmetastar {%
Sotani2007}%
\begin{APACrefauthors}%
Sotani, H.%
, Kokkotas, K\BPBI D.%
\BCBL {}\ \BBA {} Stergioulas, N.%
\end{APACrefauthors}%
\unskip\
\newblock
\APACrefYearMonthDay{2007}{}{},
\newblock
\unskip
\newblock
\APACjournalVolNumPages{Mon. Not. R. Astron. Soc.}{375}{}{261}.
\PrintBackRefs{\CurrentBib}

\bibitem [\protect \citeauthoryear {%
Steiner%
}{%
Steiner%
}{%
{\protect \APACyear {2012}}%
}]{%
Steiner2012}
\APACinsertmetastar {%
Steiner2012}%
\begin{APACrefauthors}%
Steiner, A\BPBI W.%
\end{APACrefauthors}%
\unskip\
\newblock
\APACrefYearMonthDay{2012}{}{},
\newblock
\unskip
\newblock
\APACjournalVolNumPages{Phys. Rev. C}{85}{}{055804}.
\PrintBackRefs{\CurrentBib}

\end{thebibliography}

\end{document}